\documentclass[pra,twocolumn,showpacs,superscriptaddress,amsmath,amssymb]{revtex4}

\usepackage{graphicx}% Include figure files

\begin{document}

\title{Spontaneous dressed-state polarization in the strong driving regime of cavity QED}

\author{Michael~A.~Armen}
\affiliation{Edward L.\ Ginzton Laboratory, Stanford University, Stanford CA 94305, USA}
\affiliation{Physical Measurement and Control 266-33, California Institute of Technology, Pasadena CA 91125, USA}

\author{Anthony~E.~Miller}
\affiliation{Edward L.\ Ginzton Laboratory, Stanford University, Stanford CA 94305, USA}

\author{Hideo~Mabuchi}%\email{hmabuchi@stanford.edu}
\affiliation{Edward L.\ Ginzton Laboratory, Stanford University, Stanford CA 94305, USA}

\date{\today}
\pacs{42.50.Pq,42.50.Lc,42.65.Pc,42.79.Ta}

\begin{abstract}
We utilize high-bandwidth phase quadrature homodyne measurement of the light transmitted through a Fabry-Perot cavity, driven strongly and on resonance, to detect excess phase noise induced by a single intracavity atom. We analyze the correlation properties and driving-strength dependence of the atom-induced phase noise to establish that it corresponds to the long-predicted phenomenon of spontaneous dressed-state polarization. Our experiment thus provides a demonstration of cavity quantum electrodynamics in the strong driving regime, in which one atom interacts strongly with a many-photon cavity field to produce novel quantum stochastic behavior.
\end{abstract}

\maketitle
%
%\begin{figure}[tb]
%\begin{center}
%\includegraphics[width=3in]{Parity_2.eps}
%%\vspace{-3mm}
%\caption{\label{Fig 2Cavity comparison} ...}
%\end{center}
%\vspace{-0.1in}
%\end{figure}

\noindent Current research in single-atom optical cavity quantum electrodynamics (cavity QED)~\cite{Mabu02} largely emphasizes the input-output properties of strongly coupled systems~\cite{Kimb98}, from normal-mode splitting~\cite{Thom92} to photon blockade~\cite{Birn05,Daya08}. While theory has predicted a wide range of quantum nonlinear-optical phenomena in the strong driving regime~\cite{Arme06}, experiments have with few exceptions~\cite{Saue04,Gupt07,Schu08,Brun96} focused on relatively weak driving conditions with average intracavity photon number $\lesssim 1$. Here we report the observation of characteristic high-bandwidth phase noise in the light transmitted through a resonantly driven Fabry-Perot cavity containing one strongly coupled atom and $10$--$100$ photons, confirming long-standing predictions of a phenomenon known as single-atom phase bistability~\cite{Kili91} or spontaneous dressed-state polarization~\cite{Alsi91}. Our results extend cavity-QED studies of quantum stochastic processes beyond the few-photon regime, open the door to experiments on feedback control of dressed-state cascades~\cite{Rein03}, and highlight the relevance of cavity QED in the development of ultra-low power nonlinear optical signal processing.

The Jaynes-Cummings model~\cite{Jayn63}, in which a two-level atom is coupled to a single-mode electromagnetic field, provides robust intuitions regarding the input-output behavior of real atom-cavity systems that incorporate multiple atomic sub-states and are subject to dissipative processes such as cavity decay and atomic spontaneous emission. It is well known that the spectroscopic normal-mode splitting of strongly coupled atom-cavity systems~\cite{Thom92,Boca04} simply reflects the eigenvalues of the lowest Jaynes-Cummings excited states, with peak widths determined by the cavity and atomic decay rates. It is equally true but less appreciated that the Jaynes-Cummings model provides crucial guidance for understanding cavity QED in the strong driving regime, in which the number of atom-cavity excitations grows $\gg 1$ and the model begins to resemble the dressed-state picture used to analyze radiative processes in free space. A clear example arises with the phenomenon of spontaneous dressed-state polarization as described by Alsing and Carmichael~\cite{Alsi91}, in which the highly-excited Jaynes-Cummings model is seen to `factor' into two quasi-harmonic ladders of states. The atom-cavity system tends to localize transiently on one sub-ladder or the other, resulting in one of two different phase shifts of the transmitted light, with stochastic switching between sub-ladders induced by individual atomic spontaneous emission events. We have detected and characterized atom-induced phase fluctuations of this type and achieve quantitative agreement with predictions based on an elementary cavity QED model incorporating Jaynes-Cummings dynamics, dissipation and external driving.

We employ a standard cavity QED apparatus~\cite{Mabu96} in which a cloud of laser-cooled $^{133}$Cs atoms is dropped over a high-finesse Fabry-Perot cavity (cavity length $l\approx 72$ $\mu$m, field decay rate $\kappa/2\pi\approx 8$ MHz); the cavity resonance frequency is fixed relative to the atomic $(6S_{1/2},F=4)\rightarrow (6P_{3/2},F=5)$ transition (dipole decay rate $\gamma_\perp/2\pi\approx 2.6$ MHz) using an auxiliary laser~\cite{Mabu99}. The cavity transmission is monitored using a laser probe and balanced homodyne detector. The cloud position and density can be adjusted so that isolated transits of individual atoms through the cavity mode volume are observed in the transmission signal. Atoms free-fall through the cavity but generally are subject to both optical pumping and forces induced by the cavity field. As the cavity mode forms a standing wave with Gaussian transverse profile (waist $\approx 28.5$ $\mu$m), the strength $g$ of the coherent atom-cavity coupling is a function of the atomic position (with maximum value $\approx 16$ MHz in our setup). We select transit events in which optical pumping and the atomic trajectory lead to maximal values of $g$ by initializing the cavity probe in a detuned configuration that produces a real-time photocurrent directly related to $g$. When a set threshold is reached during a single-atom transit, the probe frequency and power are quickly shifted to desired values for data acquisition. Using this triggering scheme we obtain phase-quadrature homodyne data in which near-maximal atom-cavity coupling strength is apparently maintained for up to $\sim 50$ $\mu$s, limited by optical pumping into the dark $(6S_{1/2},F=3)$ hyperfine state.

\begin{figure}[tb]
\begin{center}
\includegraphics[width=3.4in]{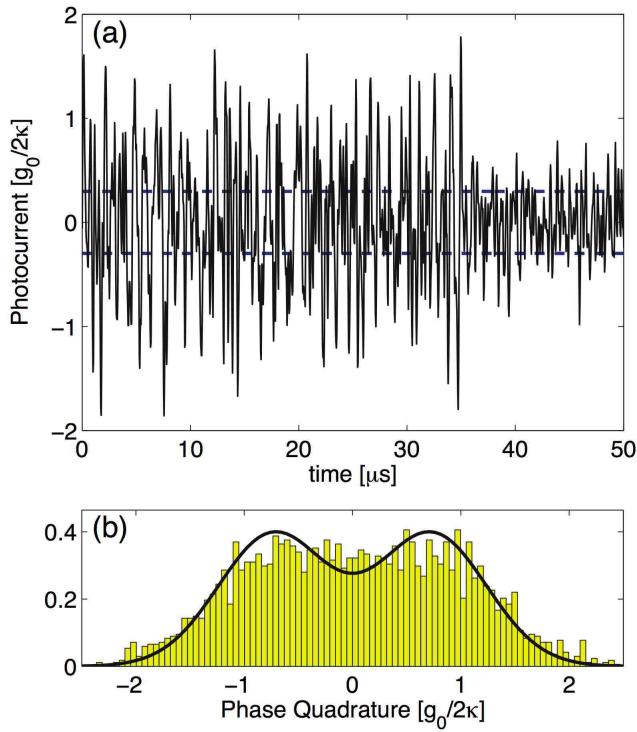}
%\vspace{-3mm}
\caption{\label{fig:trajec} (a) Black solid trace is a photocurrent segment recorded with input power such that $N\approx 20$, filtered to a bandwidth of 4 MHz. Blue dashed horizontal lines indicate the standard deviation of the optical shot noise. The atom is optically pumped to a dark state at time $t\sim 35$ $\mu$s, resulting in an abrupt disappearance of the atom-induced excess phase noise. Units are referred to the phase quadrature amplitude of the intra-cavity field. (b) Histogram of filtered photocurrent segments (0.1--8 MHz bandpass) from multiple atom transits; see text for explanation of the theoretical curve.}
\end{center}
\vspace{-0.1in}
\end{figure}

%\begin{figure}[tb]
%\begin{center}
%\includegraphics[width=3.4in]{Fig2.eps}
%%\vspace{-3mm}
%\caption{\label{fig:bimodal} (a) Histogram of filtered photocurrent segments (0.1--8 MHz bandpass) from multiple atom transits; see text for explanation of the theoretical curve. (b) Reconstruction of a two-state switching trajectory (red) from a segment of the homodyne photocurrent (black, 10 MHz bandwidth, $N\approx 37$) produced via posterior decoding~\cite{Durb98}. Note that at time $t\sim 43$ $\mu$s the reconstruction algorithm correctly identifies the end of the atom-induced fluctuations and infers a `dark' signal state with zero mean and Gaussian shot-noise fluctuations only.}
%\end{center}
%\vspace{-0.1in}
%\end{figure}

Fig.~\ref{fig:trajec}a shows a representative example of the single-shot photocurrents thus obtained. A distinct transition in the signal can be seen at time $t^*\sim 35$ $\mu$s. The photocurrent variance for $t>t^*$ corresponds to optical shot noise while for $t<t^*$ the variance is clearly larger, indicating a significant effect of the atom on light transmitted through the cavity. We interpret the change as an optical pumping event in which the atom is transferred to the dark hyperfine ground state, and have verified that such events can be suppressed by adding an intracavity repumping beam. Although the signal-to-noise ratio in our measurements is limited, a histogram of photocurrent segments from multiple atom transits (Fig.~\ref{fig:trajec}b) reveals a flat-topped distribution supporting the theoretical expectation of random-telegraph (rather than Gaussian) statistics of the atom-induced phase noise. The smooth curve is a theoretical prediction produced by fitting the sum of two Gaussian functions (constrained to have width corresponding to optical shot noise) to histograms generated via quantum trajectory simulations of our cavity QED model~\cite{Carm93,Mabu98}. In the limit of low sampling noise our experimental histogram would be expected more clearly to display such a bimodal distribution, although some deviations from ideal theory would presumably emerge because of residual atomic motion and optical pumping among Zeeman sub-states.

\begin{figure}[tb]
\begin{center}
\includegraphics[width=3.4in]{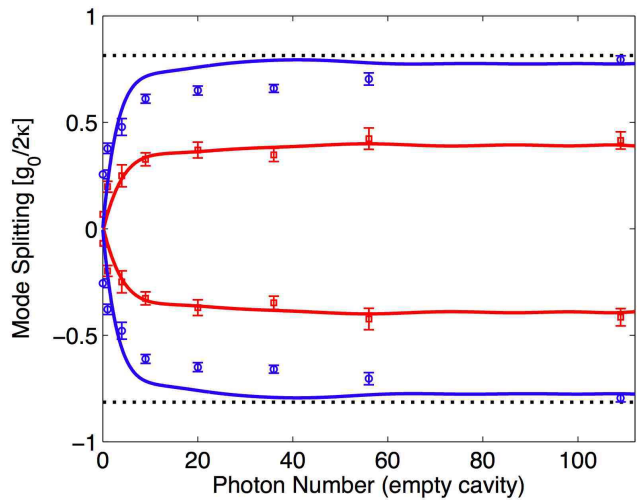}
%\vspace{-3mm}
\caption{\label{fig:splittings} Driving-strength dependence of the splitting (in units of phase quadrature amplitude of the intra-cavity field) between the centroids of bi-Gaussian fits to photocurrent histograms as in Fig.~\ref{fig:trajec}b. The experimental data (points with error bars) are directly compared with theoretical predictions based on quantum trajectory simulations (solid curves) and the cavity QED Master Equation (see text). Blue points and curves are computed with data and simulations filtered to 10 MHz; red points and curves reflect filtering at 2 MHz, which results in a decrease in apparent switching amplitude since 2 MHz is well below the purported switching frequency. Black horizontal dotted lines indicate the splitting predicted by steady-state solution of the Master Equation for large $N$.}
\end{center}
\vspace{-0.1in}
\end{figure}

In Fig.~\ref{fig:splittings} we summarize a comparison of experimental and theoretical photocurrent histograms, such as the one depicted in Fig.~\ref{fig:trajec}b, across a range of probe powers. The probe power is conventionally parametrized by the mean intracavity photon number $N$ that would be produced in an empty cavity; note that for $N\gtrsim 5$ the mean intracavity photon number produced with a strongly coupled atom is $\sim N-1$. We display the best-fit values of the two Gaussian centroids obtained in fits to data and simulations; the blue points and theoretical curves are for the maximal signal bandwidth of 10 MHz while the red points and curves are computed with signals and simulations filtered to 2 MHz. The horizontal lines indicate the splitting predicted by steady-state solution of the cavity QED Master Equation in the asymptotic region of large $N$. Our data closely match the predicted sharp onset of atom-induced phase fluctuations for lower values of $N$ and also asymptote correctly for high $N$. That the splitting becomes independent of $N$ for high $N$ is a distinctive feature of spontaneous dressed-state polarization~\cite{Alsi91}.

\begin{figure}[tb]
\begin{center}
\includegraphics[width=3.4in]{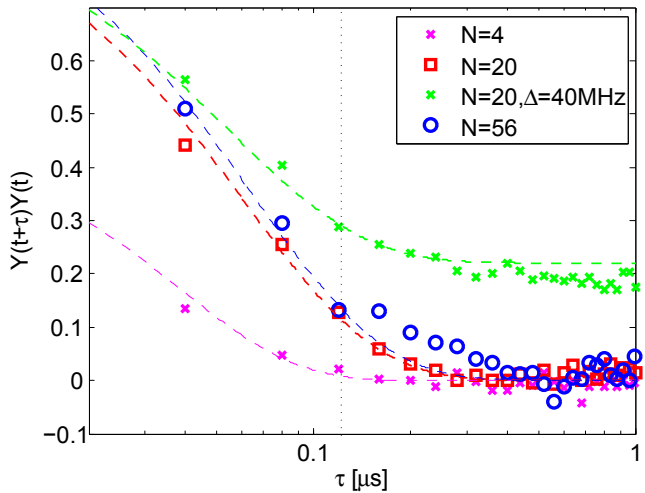}
%\vspace{-3mm}
\caption{\label{fig:autocorr} Autocorrelation functions of the photocurrent $Y(t)$ (units as in Figs.~\ref{fig:trajec},~\ref{fig:splittings}) obtained under a range of driving parameters. Experimental autocorrelations are computed after ac-filtering the photocurrents (20 kHz cutoff) to suppress artifacts caused by atomic motion and optical pumping. Points are experimental data and curves are theoretical predictions based on the cavity QED Master Equation for four different parameter sets (see legend and text).}
\end{center}
\vspace{-0.1in}
\end{figure}

In Fig.~\ref{fig:autocorr} we display the autocorrelation functions of experimental photocurrent segments for four characteristic sets of probe parameters, together with theoretical predictions. Data points and theoretical curves are displayed for $N=(4,20,56)$ with the atom, cavity and probe frequencies all coincident, and for $N=20$ with the cavity and probe frequencies coincident but 40 MHz below the atomic resonance frequency. The vertical line indicates the predicted average period ($2/\gamma_\perp$) of spontaneous phase-switching events. In the resonant cases, the rapid growth with $N$ of predominantly short-timescale photocurrent fluctuations agrees with theory and reinforces our identification of the atom-induced phase noise with switching caused by spontaneous emission. The increased correlation timescale for the detuned case follows from a tendency of the atom-cavity system to favor one of the sub-ladders discussed by Alsing and Carmichael~\cite{Alsi91}. We should note that theory predicts identical autocorrelation functions for atom-cavity detuning of either positive or negative sign, but in the experiment setting $\Delta<0$ results in a drastic reduction of atom transit signals as a result of repulsive mechanical forces exerted by the detuned intracavity light.

\begin{figure}[tb]
\begin{center}
\includegraphics[width=3.4in]{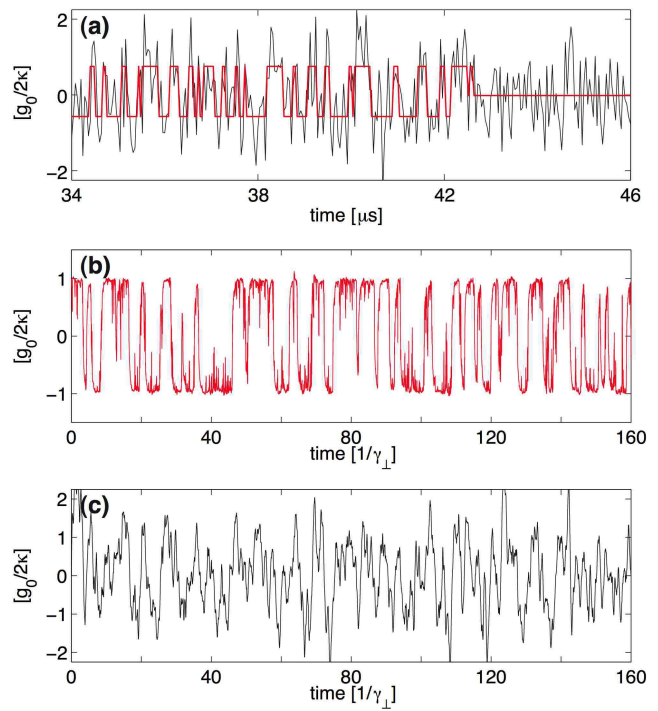}
%\vspace{-3mm}
\caption{\label{fig:bimodal} (a) Theoretical reconstruction (red) produced via posterior decoding~\cite{Durb98} of a two-state switching trajectory from a segment of the experimental homodyne photocurrent (black, 10 MHz bandwidth, $N\approx 37$). Note that at time $t\sim 43$ $\mu$s the reconstruction algorithm correctly identifies the end of the atom-induced fluctuations and infers a `dark' signal state with zero mean and Gaussian shot-noise fluctuations only. (b) Quantum trajectory simulation of the expected switching behavior in a cavity QED system (conditional expectation value of the phase quadrature amplitude of the {\em intra}-cavity field) using parameter values of the current experiment. (c) Simulated phase-quadrature homodyne photocurrent corresponding to (b), including shot-noise and finite bandwidth as in the experimental data. Note that the duration of the simulations in (b) and (c) are $\approx 10$ $\mu$s.}
\end{center}
\vspace{-0.1in}
\end{figure}

Although statistical comparisons strongly support the conclusion that our data reflect atom-induced phase noise associated with spontaneous dressed-state polarization, direct visualization of `bistable' switching in our measured photocurrents is difficult because of the modest ratios $g:\kappa:\gamma_\perp$ realized in our experiment. We have been limited in this regard by the properties of the mirrors we had available at the time the experiment was assembled (significant improvements should be possible with commercial mirror technology~\cite{Mabu98}), and by a pronounced birefringence of the assembled cavity that forced us to use linear rather than circular probe polarization~\cite{Birn05b}. Furthermore our maximum digitization rate in recording the photocurrent was $2.5\times 10^7$ samples per second, whereas the switching rate induced by atomic spontaneous emission should be $\gamma_\perp/2\approx 8$ MHz. It is nevertheless possible to utilize standard techniques for hidden Markov models (HMM's) to attempt to reconstruct two-state switching trajectories from individual photocurrent records. In Fig.~\ref{fig:bimodal} the red trace shows the result of applying a standard posterior decoding algorithm~\cite{Durb98} to the photocurrent shown in black, assuming a simple three-state HMM in which the states corresponds to negative, positive, and zero values of the conditional expectation of the intra-cavity phase quadrature amplitude~\cite{PW09}. The red trace schematically indicates, at each point in time, which of the three states has highest posterior probability. Although we are unable directly to check the accuracy of this reconstruction, numerical experiments based on simulations such as the one shown in Figs.~\ref{fig:bimodal}b,c predict an accuracy $\sim 10\%$.

The demonstration of spontaneous dressed-state polarization constitutes an important step in the study of nonlinear optical phenomena at ultra-low energy scales, where quantum fluctuations interact with nonlinear mean-field dynamics to generate complex stochastic behavior~\cite{Arme06}. Single-atom cavity QED provides an important setting for such studies, as theoretical models that accurately capture the behavior of real physical systems can be formulated with sufficiently low variable count to enable direct numerical simulations for conceptual analysis and comparisons with experiment. These models, which in some cases are amenable to systematic reduction techniques~\cite{VanH06,Niel09}, provide a unique resource for research in non-equilibrium quantum statistical mechanics and on quantum--classical correspondence in nonlinear dynamical systems~\cite{Arme06}.

Cavity QED in the strong driving regime may also be of interest for exploratory studies in attojoule nonlinear optics. While the difficulties of all-optical information processing are well known~\cite{Smit84,Mill90}, the prospect of photonic interconnect layers within microprocessors~\cite{Beau07} is reviving interest in ultra-low energy optical switching. The characteristic scale of 50 photons corresponds to an energy $\approx 12$ aJ, representing an operating regime for optical switching that lies significantly below foreseeable improvements in existing technology but stays above the single-photon level where propagation losses and signal regeneration would seem to be dominant engineering concerns. Ultra-low energy nonlinear optical effects should be achievable with nanophotonic implementations of cavity QED~\cite{Srin07,Fara08}, providing a potential path toward large-scale integration. Even with our modest values of $g$ and $\kappa$, the atom-induced phase shift of light transmitted through the cavity is $\pm 0.15$ rad for input power corresponding to $N=20$. Each atomic spontaneous emission event that switches the phase of the transmitted light dissipates a mere $0.23$ aJ of energy, while the light transmitted through the cavity during a typical interval between switching events carries an optical signal energy $\approx 3.3$ aJ at $N=20$. Such a 1:10 ratio between the switching energy and the energy of the controlled signal in nonlinear-optical phase modulation would be highly desirable for the implementation of cascadable photonic logic devices, and is not generally achieved in schemes based on single-photon saturation effects in cavity QED~\cite{Turc95,Engl07}. While it is certainly not clear whether the specific phenomenon of spontaneous dressed-state polarization can be exploited in the design of practicable switching devices, we hope that our demonstration will serve to draw some attention up from the bottom of the Jaynes-Cummings ladder towards the strong-driving regime of cavity QED.

This research has been supported by the NSF under PHY-0354964, by the ONR under N00014-05-1-0420 and by the ARO under W911NF-09-1-0045. AEM acknowledges the support of a Hertz Fellowship. We thank David Miller for enlightening conversations and Dmitri Pavlichin for crucial technical assistance in the preparation of Fig.~\ref{fig:bimodal}a.

\end{document}